\documentstyle[prb,aps,twocolumn,floats]{revtex}
\input epsf
\begin{document}
\twocolumn[\hsize\textwidth\columnwidth\hsize\csname
@twocolumnfalse\endcsname
\title{A Current Induced Transition in atomic-sized contacts of
metallic alloys}
\author{Jan W.T. Heemskerk$^1$, Yves Noat$^{1,2}$, David
J. Bakker$^1$, and Jan M. van Ruitenbeek$^1$}
\address{$^1$Kamerlingh Onnes Laboratorium, Universiteit Leiden, Postbus
9504, 2300 RA Leiden, The Netherlands}
\address{$^2$Groupe de Physique des Solides, Campus Jussieu, Tour 23, 2 Place
Jussieu, 75251 Paris Cedex 05, France}
\author{Barend J. Thijsse, Peter Klaver}
\address{Laboratory of Materials Science, Delft University of Technology,
Rotterdamseweg 137, 2628 AL Delft, The Netherlands}
\date{\today}
\maketitle

\begin{abstract}

We have measured conductance histograms of atomic point contacts
made from the noble-transition metal alloys CuNi, AgPd, and AuPt
for a concentration ratio of 1:1. For all alloys these histograms
at low bias voltage (below 300 mV) resemble those of the noble
metals whereas at high bias (above 300 mV) they resemble those of
the transition metals. We interpret this effect as a change in the
composition of the point contact with bias voltage. We discuss
possible explanations in terms of  electromigration and
differential diffusion induced by current heating.

\end{abstract}

\vskip2pc] \narrowtext
\section{Introduction\label{Introduction}}

The scale of electronic devices will soon be reduced to a level
where material properties will be very different from the bulk,
necessitating research into very small scale devices. A metallic
point contact is perhaps the most simple example of an
atomic-sized device. It consists of a connection of a few atoms,
typically between 1 and 1000, between two macroscopic electrodes.
The size of the contact is therefore of the same order of
magnitude as the Fermi wavelength of the electrons. In this limit,
the conductance is related to the transmission of an electron wave
through the system. For a free electron gas, the transmission only
depends on the size of the constriction. As a result, the
conductance in a 2-dimensional electron gas has been seen to
increase by steps as a function of the width of the
constriction\cite{Wees1,Wharam1}. The situation is more
complicated in a metallic contact, where the transmission depends
on the geometry of the contact as well as on the atomic structure
of the atoms forming it. In the simplest case the contact consists
of a single atom, where the conductance has been shown to be
determined by the valence orbitals of the atom forming the
contact\cite{Scheer1,Cuevas1}.

Several ways exist to create a point contact. The method used in
this research is the so-called mechanically controllable break
junction (MCBJ) technique\cite{Muller1}. Its principle is very
simple; it consists of breaking a metallic wire thereby creating a
clean fracture surface. After rupture, an atomic-sized contact can
be made by indenting the broken ends of the wire into each other.
This contact is subjected to a repeated cycle of breaking and
indenting, during which the conductance of the contact is
measured. As the contact is elongated, the diameter of the
constriction reduces and consequently the conductance decreases.

For most metals at small contacts, the conductance decreases in
steps of the order of $G_0 = { 2e^2 / h}$, the quantum of
conductance. It has been shown that these steps are due to
rearrangements of atoms in the contact\cite{Rubio1}. The last
plateau is generally assumed to correspond to a single atom
contact\cite{Krans1}.

The electrical and mechanical properties of atomic point contacts
made from pure metals have been extensively
studied\cite{Olesen1,Costa-Kramer1,Krans2,Hansen1}. For noble
metals (Cu, Ag, Au) the last plateau of a conductance trace,
corresponding to a contact of one atom, has a conductance value
around 1\,$G_0$. On the other hand, transition metals (of which we
will consider Ni, Pd, Pt) have a smallest conductance value around
1.5\,$G_0$, or higher. These values are explained by the fact that
noble metal atoms have a single $s$ valence orbital and transition
metal atoms have five $d$ valence orbitals in addition. Very few
studies have been made of the properties of {\em alloys} at the
atomic scale.

Recently, point contact studies were made of random alloys of a
transition metal and a noble metal, namely gold and
paladium\cite{Enomoto1} and copper and nickel\cite{Bakker1}, for
different concentration ratios. For AuPd, the addition of Pd only
leads to a decrease in height of the Au conductance peak, and no
shift is observed. We propose below a different explanation for the
persistence of the Au peak up to a high concentration of Pt or Pd.
The CuNi study was aimed at an investigation of the influence of
impurity scatterers on the point contact conductance. For low Ni
concentrations the nickel atoms can be
considered as impurity scatterers for the electrons, that lead to a
decrease of the conductance resulting in a shift towards lower values
of the peaks in the conductance histogram. This shift grows linearly
with the Ni concentration. For higher concentrations of transition
metal atoms, it
is difficult to predict how the contact will behave and what effects
might occur. Naively, one would expect a single-atom contact to be
randomly formed by either a noble (Cu, Au) or a transition metal (Ni,
Pd) atom, with a probability depending on concentration. However, this
does not take into account several effects that could influence the
formation of the contact, such as segregation, diffusion and
possibly electromigration. Indeed, here we show that this simple
picture does not hold. In addition we report an unexpected
transition in the conductance characteristics as a function of the
bias voltage for random transition-noble metal alloys.
Complementary to the experiments we have performed molecular
dynamics simulations in order to obtain information on the
structure and composition of the atomic point contacts that is not
provided by a conductance measurement.

\section{Experiment\label{Experiment}}

The alloys that were used in this experiment, apart from
copper-nickel, were silver-paladium and gold-platinum, all at a
1:1 concentration ratio. These materials were chosen for the
following reasons: First, they are miscible at any concentration.
Second, the conductance of the noble metal atoms (Cu, Ag, Au) is
different from that of the transition metal atoms (Ni, Pd, Pt),
(see above) which allows us, in principle, to distinguish
single-atom contacts of noble metals and those of transition metal
atoms.

Samples were made by arc-melting equal amounts of Cu (Ag, Au) and
Ni (Pd, Pt) and quenching to room temperature. CuNi and AgPd form
random alloys. Below $\sim$1260\,C, AuPt may segregate, but we see
little evidence for segregation in our data. Of these samples
wires are made and annealed for several hours at 900\,C. A 1.5\,cm
long piece of wire is then glued on a phosphor-bronze substrate by
two drops of epoxy on both sides of a notch made in the middle of
the wire. The wire is broken at low temperature (4.2\,K) by
applying a mechanical force on the substrate. The two broken ends
are then brought back together and the bending is controlled by
means of a piezo element allowing us to obtain a contact of any
size, where the contact elongation can be controlled with an
accuracy better than 0.01\,\AA. The conductance is measured in a
four point configuration.

Individual conductance traces (plots of the conductance vs the
elongation of the wire) differ from each other because the
positions of the atoms are different each time a new contact is
formed. For a more quantitative view of the conductance of the
contact a histogram technique is used that will indicate preferred
values of the conductance when the size of the contact decreases.
A histogram is constructed as follows: the $y$-axis of an indivual
conductance trace is divided into a number of bins and we count
the number of data-points that fall in those bins. Peaks in the
histogram represent preferred values of the conductance.

Histograms were created from $\sim$2500 individual conductance
traces. Between these individual traces the broken ends of the
wire are indented to a large contact size that is set by a
predetermined  value for the conductance. The usual indentation
depth was to a conductance of 25--30\,$G_0$, corresponding to a
contact of approximately as many atoms.

For a further investigation of the noble-transition metal point
contacts, classical molecular dynamics simulations have been
performed on a CuNi point contact. The system was simulated using
two-atom potentials and a Johnson-Oh embedding function
formalism\cite{Johnson-Oh}. The motion of the atoms in a point
contact of a random alloy at low temperature was calculated while
the contact was being stretched.

\section{Results\label{Results}}

In Fig.\,\ref{Fig.AuPthistoup} histograms of AuPt recorded with
four different voltages applied across the contact (bias voltage)
are shown. All histograms are normalized by the area under the
curves in the interval [0,10]\,$G_0$.The histogram at low bias
voltage exhibits a strong peak positioned somewhat below 1\,$G_0$.
The amplitude of the peak initially decreases slightly as the bias
voltage is increased but at $V_{\rm bias} =$ 400\,mV it has
disappeared and instead a peak positioned at G$\simeq$1.9\,$G_0$
appears. It is important to notice that this peak is located
approximately at the position of the first peak of pure Pt.

Measurements of the conductance of copper-nickel and
silver-paladium showed that also for these alloys the low bias
voltage histogram resembles that of the noble metal having a
dominant peak slightly below 1\,$G_0$, while at high bias it
resembles the conductance histogram of the transition metal and
all the weight below 1.5\,$G_0$ has disappeared.

\begin{figure}[!t]
\begin{center}
 \leavevmode
 \epsfxsize=70mm
 \epsfbox{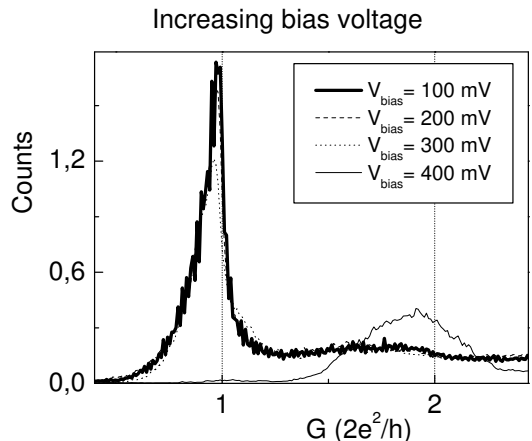}
 \caption{\label{Fig.AuPthistoup}
Normalized histogram for AuPt for different bias voltages,
starting from 100\,mV and increasing to 400\,mV. The position of
the peak changes from 0.95\,$G_0$ to 1.9\,$G_0$ when the bias
 voltage is increased from 300 to 400\,mV. The indentation
of the contact after each scan was to a conductance of 30\,$G_0$.}
\end{center}
\end{figure}

\begin{figure}[!t]
\begin{center}
 \leavevmode
 \epsfxsize=70mm
 \epsfbox{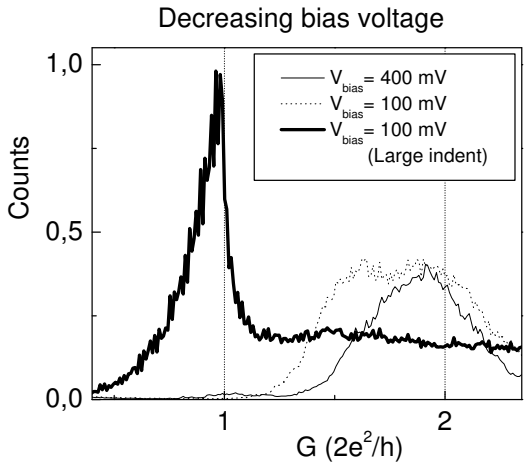}
  \caption{\label{Fig.AuPthistodown}
Histogram of AuPt when decreasing the bias voltage for different
values above and below the critical value for inducing the
transition. The peak around 1\,$G_0$ does not reappear unless a
large contact is made. The indentation of the contact after each
scan is to a conductance of 30\,$G_0$.}
\end{center}
\end{figure}

After a high bias voltage measurement, decreasing the voltage
across the contact did not result in a return to a
noble-metal-like histogram, as can be seen in
Fig.\,\ref{Fig.AuPthistodown}. For all alloys, only after full
indentation of both ends of the wire (by putting the voltage on
the piezo to zero), creating a contact of mesoscopic size ($>$
1000 atoms), a noble-metal-like histogram was recovered.

\begin{figure}[!b]
\begin{center}
 \leavevmode
 \epsfxsize=70mm
 \epsfbox{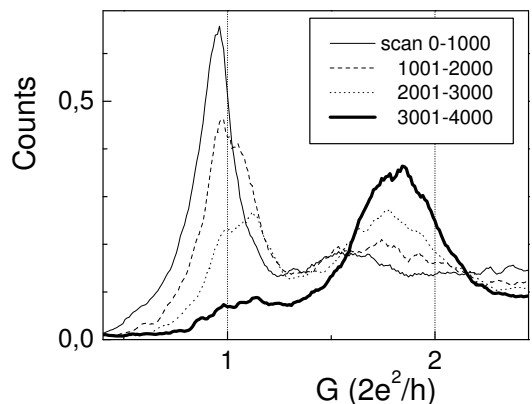}
  \caption{\label{Fig.AuPttimedep}Time dependence of the conductance
  of an AuPt point contact, starting from a low-bias structure. One
 can observe the gradual change in the conductance peak from $\sim$0.95 to
 $\sim$1.8\,$G_0$. The bias voltage is 375\,mV, the indentation of the
 contact after each scan is to a conductance of 30\,$G_0$. The time
 elapsed between scan 1 and scan 4000 is approximately 1500\,s.}
\end{center}
\end{figure}

From a conductance histogram it is impossible to tell whether the
shift in conductance is gradual or sudden since the histogram is
accumulated from many conductance scans. Therefore we decided to
measure a number of subsequent histograms of a smaller number of
scans (500--1000) to be able to observe possible time-evolution of
the contact. Again, the results are similar for the three alloys.
At intermediate bias voltage (between 300 and 400\,mV) it is
possible to observe the change from a noble to
transition-metal-like conductance histogram over the course of a
small number of histograms as can be seen in
Fig.\,\ref{Fig.AuPttimedep}.

\begin{figure}[!t]
\begin{center}
 \leavevmode
 \epsfxsize=70mm
 \epsfbox{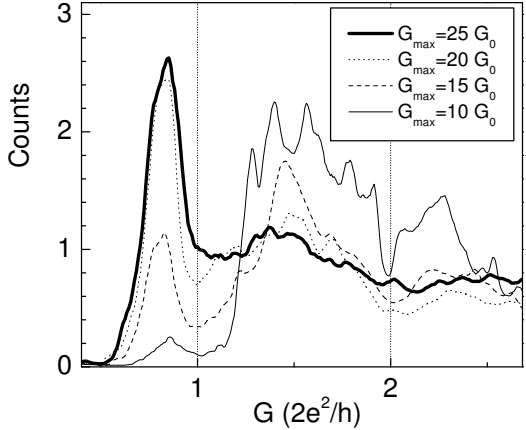}
  \caption{\label{Fig.AgPdindent}
Histogram for a AgPd point contact for different indentation
depths at $V_{\rm bias}$=\,220 mV. For each histogram we start
from a contact having a fresh low-bias characteristic structure.
One can observe that the transition occurs faster at smaller
indentation.}
\end{center}
\end{figure}

We have also studied the influence of the indentation depth of the
broken ends of the wires on the transition in the contact. In this
experiment, the largest contact size between individual traces of
a histogram was varied for a number of histograms. This contact
size between traces is a measure for the indentation depth, and is
characterized by a certain conductance value. It is observed
(Fig.\,\ref{Fig.AgPdindent}) that at smaller indentation depth the
change from noble to transition-metal-like appears sooner (both at
lower bias voltage and in a shorter time period) than for larger
indentation. The additional fine structure in the case of very
small indentation (10\,$G_0$) in Fig.\,\ref{Fig.AgPdindent} can be
explained by the fact that at small indentation the number of
possible atomic configurations is reduced compared to a large
indentation, and consequently the number of conductance values is
limited. At very low indentation depths this can result in a
repetition of a limited number of evolution paths for the contact.

In order to determine if the transition is related to the
direction of the current in the contact we have conducted the same
experiment using an ac voltage instead of a dc one. The frequency
of the bias voltage was $\omega =$10\,kHz, which was limited by
the frequency transfer characteristics of our setup. The current
and voltage over the sample were detected using two lock-in
amplifiers. We found that the results obtained from the ac-voltage
experiment are similar to ones from the dc-experiments. We
conclude that the transition is not related to the direction of
the current, at least for time scales longer than 0.1\,ms.

In the molecular dynamics simulations, a random alloy of copper
and nickel atoms was created, shaped like an hourglass, which was
subjected to about 30 cycles of indenting and breaking of the
contact at a temperature of 4K. The main outcome of the
simulations is that the randomness of the alloy is preserved; no
segregation was observed for up to 18 cycles of
indentation to a contact size of 10--50 atoms and
subsequent breaking. Our model does not allow us to simulate an
electrical potential across the contact or a current through it.
Another significant drawback is the fact that even though the
timescale of the simulation was $\sim 10^{-9}$\,s, which is 8
orders of magnitude smaller than in the experiment, the
calculation time was one or two days for a single cycle, thus
limiting the number of simulations we could perform.

\section{Discussion\label{Discussion}}

\begin{figure}[!t]
\begin{center}
 \leavevmode
 \epsfxsize=70mm
 \epsfbox{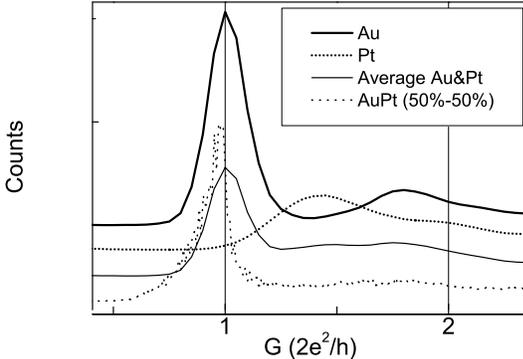}
  \caption{\label{Fig.AuPtaverage}
Histograms of pure gold, pure platinum, their average and a histogram 
of gold-platinum measured at low bias voltage. The graphs are shifted
on the y-axis for clarity.}
\end{center}
\end{figure}

As is illustrated for Au and Pt in Fig.\,\ref{Fig.AuPtaverage}, at
low bias voltage the histograms of the alloys resemble the
histograms of the noble metals (Cu, Ag, Au) rather than those of
the transition metals (Ni, Pd, Pt), with a distinct peak slightly
below 1\,$G_0$, but no peak around $\sim$1.5--1.8\,$G_0$. However,
when a high bias voltage is applied to the contact the picture is
reversed. In fact, for bias voltages above the threshold value the
conductance histogram only exhibits a peak located at a
conductance value of 1.5--1.8 $G_0$, and is therefore similar to a
transition metal histogram.

This suggests that as a result of the bias voltage applied across
the contact the noble metal atoms are expelled from it, and only
transition metal atoms remain. The peak around 1\,$G_0$ for
low-bias voltage indicates the presence of noble metal atoms; the
fact that there is no weight in the conductance histogram below
1.5\,$G_0$ in case of a high-bias voltage indicates the absence of
noble metal atoms.

From the time dependence of the current-induced changes we
conclude that the composition of the contact is irreversibly
modified, locally. Fig.\,\ref{Fig.AuPttimedep} illustrates how
this modification gradually builds up and
Fig.\,\ref{Fig.AuPthistodown} shows that the initial properties
are only recovered after making a large indentation. These
observations can be understood by assuming that Au (Cu, Ag) atoms
are driven away from the contact at high bias, leaving a nearly
pure Pt (Ni, Pd) contact. After indentation to large contacts
fresh Au (Cu, Ag) atoms are anew mixed into the contact area. The
importance of the irreversibility of the shift with bias voltage
should not be underestimated. It indicates that the shift cannot be
attributed to a straightforward effect in pure metals, since such an
effect would be reversible.

This view is supported by the fact that the transition is also
dependent on the indentation depth. At higher indentation depths
the transition takes place at a higher bias voltage. This agrees
with the larger region that has to be depleted of noble metal
atoms for larger indentation depths in order to have a contact
consisting purely of transition metal atoms. The number of atoms
which have to be expelled therefore increases with the
indentation.

There are two principle candidates for explaining the observed
change in composition of the contacts when a bias voltage is
applied across the contact: electromigration, and differential
thermal diffusion. Electromigration is the motion of atoms in a
conductor under the influence of an applied voltage. It is
traditionally described in terms  of a direct force and wind
force\cite{Sorbello1,Schmidt1}. The direct force is believed to be
due to the electric field and the effective charge of the atom,
and the wind force is due to the scattering of the current
carrying electrons resulting in a momentum transfer. More recent
work suggests that it is not meaningful to distinguish these two
components and that one should consider the total force to be due
to an induced bias in activation barriers for migration
\cite{Hoekstra00}.

Scattering of current-carrying electrons causes Joule heating of
the contact \cite{Brom1,Todorov1,Brom2} resulting in a temperature
gradient leading to a diffusion of atoms away from the contact.
This could lead to a change in the composition of the contact when
the diffusion coefficients of the different types of atom are
unequal.
The type of atom with the largest diffusion coefficient would be
expelled from the contact. We estimate that a bias voltage of
400\,mV raises the lattice temperature locally to $\sim$400\,K
\cite{Brom1,Todorov1,Brom2}. The diffusion constants for the noble
metals are usually much larger than those for the transition
metals.

The experiments discussed above all support the view that the
composition of the contact is changed, but unfortunately they do
not shed any light on the mechanism behind it. The dependence of
electromigration on the direction of the current could possibly
allow us to distinguish between thermal diffusion and
electromigration. In an ac-experiment the direction of the
electrical force would change with the potential. When the
polarity of the potential changes rapidly the average force on the
atoms would be negligible, and only thermal diffusion would remain
as a possible cause for the transition. The result of the
ac-experiments were the same as those from the dc-experiments.
Therefore we conclude that the transition in the contact is more
likely due to the second effect. Electromigration is not fully
excluded since the gradient in the current density would drive the
atoms away from the contact area, making the return path much less
effective.

\begin{figure}[!t]
\begin{center}
 \leavevmode
 \epsfxsize=80mm
 \epsfbox{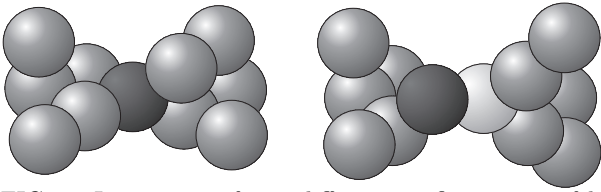}
  \caption{\label{transitiondraw}
Impression of two different configurations of last contact, formed
by one atom (dark, left) or two atoms in series (dark and light,
right).}
\end{center}
\end{figure}

The outcome of the simulations did not show a preference for the
formation of noble-metal last atom contacts over transition-metal
atom contacts. The fact that we could not simulate a voltage or
current in the contact prevents us from investigating a possible
preference for the formation of transition metal atom contacts at
high bias.  However, the simulations did shed some more light on
the dynamics of the point contact. In the discussion we have
previously assumed that the last contact would be formed by one
atom between the two shoulders of the banks. In the simulations we
observed that the last contact is often formed by two atoms, one
in each shoulder, see Fig.\ref{transitiondraw}.

This changes the question of what the conductance of a noble- or
transition-metal-atom contact is, to the question of what is the
conductance of a contact between a noble and a transition-metal
atom. Based on the work in Refs. \cite{Scheer1,Cuevas1} we
speculate that the conductance of a contact between a copper and a
nickel atom is determined by the copper atom, since it has the
smallest number of valence orbitals or conductance channels, which
would limit the conductance to a single channel with a conductance
somewhat below 1\,$G_0$. If, as the simulations suggest, most of
the final contacts are formed by two atoms in series, then
$\sim$75\% of the final contacts would have approximately the
conductance of copper. This may also explain why we cannot
distinguish a nickel conductance peak in the CuNi low bias
histogram, since one would expect only $\sim$25\% of the contacts
to be formed by two nickel atoms. Thus, the conductance histogram
of CuNi would resemble the copper histogram with a peak below 1
$G_0$. The assumption that the conductance between two atoms is
determined by the atom with the smallest number of conductance
channels remains to be further investigated. However, recent work
on a hydrogen molecule trapped between two Pt electrodes shows
that this system has a single nearly fully transmitting
conductance channel, presumably due to the single $s$ orbital of
hydrogen \cite{smit02}.

In conclusion, we have investigated atomic point contacts made
from the noble-transition metal alloys AuPt, CuNi, and AgPd at a
concentration ratio of 1:1. At low bias voltage the histograms
resemble the ones of the noble metals whereas this situation is
reversed at high bias. Our interpretation is that the contact is
depleted of noble metal atoms with high bias. The concentration
can therefore be tuned with the bias voltage. The mechanism we
propose relies on the strong local gradients in lattice temperture
and/or current density, driving the noble-metal atoms away from
the junction at high voltage bias.

We thank A. Lodder, R.H.M. Smit, C. Untiedt, and I.K. Yanson for
helpful discussions, and R.W.A. Hendrikx and M.B.S. Hesselberth
for assistance with the sample preparation. This research has been
supported by a European Community Marie Curie Fellowship under
contract number HPMF-CT-1999-00196.


\begin{thebibliography}{12}

\bibitem{Wees1}
B.J. van Wees, H. van Houten, C.W.J. Beenakker, J.G. Williamson,
L.P. Kouwenhoven, D. van der Marel, and C.T. Foxon,
Phys. Rev. Lett. {\bf 60}, 848 (1988).

\bibitem{Wharam1}
D.A. Wharam, T.J. Thornton, R. Newbury, M. Pepper, H. Ahmed,
J.E.F. Frost, D.G. Hasko, D.C. Peacock, D.A. Ritchie, and
G.A.C. Jones, J.Phys.C. {\bf 21}, L209 (1988).

\bibitem{Scheer1}
E. Scheer, N. Agra\"{\i}t, J.C. Cuevas, A. Levy Yeyati, B. Ludoph,
A. Mart\'{\i}n-Rodero, G. Rubio Bollinger, J.M. van Ruitenbeek,
and C. Urbina, Nature {\bf 394}, 154, (1998).

\bibitem{Cuevas1}
J.C. Cuevas, A. Levy Yeyati, and A. Mart\'{\i}n-Rodero, Phys. Rev.
Lett. {\bf 80}, 1066 (1998).

\bibitem{Muller1}
C.J. Muller, J.M. van Ruitenbeek, and L.J. de Jongh, Physica C
{\bf 191}, 485 (1992).

\bibitem{Rubio1}
G. Rubio, N. Agra\"{\i}t, and S. Vieira, Phys. Rev. Lett. {\bf
76}, 2302 (1996).

\bibitem{Krans1}
J.M.Krans, C.J.Muller, I.K. Yanson, Th.C.M.Govaert, R.Hesper, and
J.M. van Ruitenbeek, Phys. Rev. B {\bf 48}, 14721 (1993).

\bibitem{Olesen1}
L. Olesen, E. L{\ae}gsgaard, I. Stensgaard, F. Besenbacher,
J. Schi{\o}tz, P. Stoltze, K. W. Jacobsen, and J. K. N{\o}rskov,
Phys. Rev. Lett. {\bf 72}, 2251 (1994).

\bibitem{Costa-Kramer1}
J. L. Costa-Kr\"amer, N. Garc\'{\i}a, and H. Olin, Phys. Rev. B
{\bf 55}, 12910, (1997).

\bibitem{Krans2}
J.M. Krans, J.M. van Ruitenbeek, V.V. Fisun, I.K. Yanson, and L.J. de
Jongh, Nature {\bf 375}, 767 (1995).

\bibitem{Hansen1}
K. Hansen, E. L{\ae}gsgaard, I. Stensgaard, and F. Besenbacher,
Phys. Rev. B {\bf 56}, 2208, (1997).

\bibitem{Enomoto1}
A. Enomoto, S. Kurokawa, and A. Sakai, Phys. Rev. B {\bf 65}, 125410
(2002).

\bibitem{Bakker1}
D.J. Bakker, Y. Noat, A.I. Yanson, and J.M. van Ruitenbeek, Phys.
Rev. B {\bf 65}, 235416 (2002).

\bibitem{Johnson-Oh}
This formalism was chosen following a comparison of different Embedded
Atom Method (EAM) potentials by Morishita and De la Rubio;
K. Morishita, and T. Diaz de la Rubia, Mat. Res. Soc. Symp. Proc. {\bf
396}, 39 (1995).

\bibitem{Sorbello1}
For an overview, see, for instance, R.S. Sorbello, {\em Theory of
Electromigration}, Solid State Physics {\bf 51}, 159, (1998).

\bibitem{Schmidt1}
T. Schmidt, R. Martel, R. Sandstrom, and P. Avouris,
Appl. Phys. Lett. {\bf 73}, 2173 (1998).

\bibitem{Hoekstra00}
J. Hoekstra, A.P. Sutton, T. N. Todorov, and A.P. Horsfield, Phys.
Rev. B {\bf 62}, 8568 (2000).

\bibitem{Brom1}
H.E. van den Brom, PhD-thesis Leiden University, (2000).

\bibitem{Todorov1}
T. N. Todorov, Phil. Mag. B {\bf 77}, 965 (1998).

\bibitem{Brom2}
H.E. van den Brom, A.I. Yanson and J.M. van Ruitenbeek, Physica B {\bf
252}, 69 (1998).

\bibitem{smit02}
R.H.M. Smit, Y. Noat, C. Untiedt, N.D. Lang, M.C. van Hemert, and
J.M. van Ruitenbeek, Nature, (2002) in print.

%\bibitem{Webber1}
%P.R. Webber, and D. Chadwick, Surf. Sci. {\bf 94}, L151 (1980).
%
%\bibitem{Sakurai1}
%T. Sakurai, T. Hashizume, A. Kobayashi, A. Sakai, S. Hyodo, Y. Kuk,
%and H.W. Pickering,
%Surface segregation of Ni-Cu binary alloys studied by an atom probe,
%Phys. Rev. B {\bf 34}, 8379 (1986).

%\bibitem{Abrikosov1}
%I.A. Abrikosov, and H.L. Skriver,
%Self-consistent linear-muffin-tin-orbitals coherent-potential
%technique for bulk and surface calculations: Cu-Ni, Ag-Pd, and Au-Pt
%random alloys,
%Phys. Rev. B {\bf 47}, 16532 (1993).

%\bibitem{Swartzfager1}
%D.G. Swartzfager, and M.J. Kelley,
%Correlation of surface segregation and bulk diffusion for binary
%metal alloys,
%Phys. Lett. {\bf 76A}, 86 (1980).

\end{thebibliography}
\end{document}